\documentclass[usegraphicx,usenatbib,letterpaper]{emulateapj}

\usepackage{ulem}
\usepackage{color}
\usepackage{graphics,graphicx}
\usepackage{times} 
\usepackage{amssymb}
\usepackage{amsmath}
\usepackage{natbib}
\usepackage{url}
\usepackage{multirow}
\usepackage{ctable}
\usepackage{threeparttable}
\newif\ifAMStwofonts
\AMStwofontstrue

\def\xmm{{\it XMM-Newton}}

\def\suzaku{{\it Suzaku}}

\def\integral{{\it INTEGRAL~\/}}

\def\epicpn{{\it EPIC}{\rm-pn}}
\def\epicmos1{{\it EPIC}{\rm-MOS1~\/}}
\def\epicmos2{{\it EPIC}{\rm-MOS2 ~\/}}
\def\epicmos{{\it EPIC}{\rm-MOS}}

\def\nustar{{\it NuSTAR}}

\def\ks{\hbox{$\rm\thinspace ks$}}

\def\H0{{\rm ~km~s^{-1}~Mpc^{-1}}}

\def\kev{\hbox{\rm keV}}

\def\ergpcmsqps{\hbox{$\rm\thinspace erg~cm^{-2}~s^{-1}$}}

\def\ergps{\hbox{erg~s$^{-1}$}}

\def\msun{\hbox{$M_{\odot}$}}

\def\rchi{{$\chi^{2}_{\nu}$~\/}}

\def\heasoft{\hbox{\rm{\small HEASOFT}}}
\def\nustardas{\rm {\small NUSTARDAS}}
\def\xselect{\hbox{\rm{\small XSELECT}}}
\def\xmmselect{\hbox{\rm{\small XMMSELECT}}}
\def\ftool{\hbox{\rm{\small FTOOL}}}

\def\specgroup{\rm {\small SPECGROUP}}

\def\addascaspec{\hbox{\rm{\small ADDASCASPEC~\/}}}
\def\sas{\hbox{\rm{\small SAS~\/}}}
\def\epchain{\hbox{\rm{\small EPCHAIN}}}
\def\emchain{\hbox{\rm{\small EMCHAIN}}}

\def\rmfgen{\hbox{\rm{\small RMFGEN}}}
\def\arfgen{\hbox{\rm{\small ARFGEN}}}
\def\epiclccorr{\hbox{\rm{\small EPICLCCORR}}}

\def\addascaspec{\hbox{\rm{\small ADDASCASPEC}}}

\def\grid25{\hbox{\rm{\small GRID25}}}

\def\simpl{\rm{\small SIMPL}}

\def\tbabs{\rm{\small TBABS}}
\def\tbnew{\rm{\small TBNEW}}
\def\tbnewlink{http://pulsar.sternwarte.uni-erlangen.de/wilms/research/tbabs}
\def\diskbb{\rm{\small DISKBB}}
\def\diskpbb{\rm{\small DISKPBB}}

\def\diskpn{\rm{\small DISKPN}}

\def\reflionx{\rm{\small REFLIONX}}

\def\relconv{\rm{\small RELCONV}}

\def\comptt{\rm {\small COMPTT}}

\def\ka{K\,$\alpha$}

\def\eg{{\it e.g.~\/}}

\def\ie{{\it i.e.~\/}}

\def\la{\mathrel{\hbox{\rlap{\hbox{\lower4pt\hbox{$\sim$}}}{\raise2pt\hbox{$<$}}}}}
\def\ga{\mathrel{\hbox{\rlap{\hbox{\lower4pt\hbox{$\sim$}}}{\raise2pt\hbox{$>$}}}}}

\def\d25{D$_{25}$}

\def\los{line-of-sight\thinspace}
\def\.25{0.25 keV\thinspace}

\def\fcol{\rm $f_{\rm col}$}

\def\mbh{\rm $M_{\rm BH}$}

\def\fvar{$F_{\rm var}$}

\def\hoix{\rm Holmberg\,IX X-1}

\shorttitle{Broadband X-ray Spectra of \hoix}
\shortauthors{D.~J. Walton et al.}

\begin{document}

\title{Broadband X-ray Spectra of the Ultraluminous X-ray Source
Holmberg\,IX X-1 observed with \nustar, \textit{XMM-Newton} and
\textit{Suzaku}}

\author{D. J. Walton\altaffilmark{1} \thanks{E-mail: dwalton@srl.caltech.edu},
F. A. Harrison\altaffilmark{1},
B. W. Grefenstette\altaffilmark{1},
J. M. Miller\altaffilmark{2},
M. Bachetti\altaffilmark{3,4},
D. Barret\altaffilmark{3,4},
S. E. Boggs\altaffilmark{5},
F. E. Christensen\altaffilmark{6},
W. W. Craig\altaffilmark{5},
A. C. Fabian\altaffilmark{7},
F. Fuerst\altaffilmark{1},
C. J. Hailey\altaffilmark{8},
K. K. Madsen\altaffilmark{1},
M. L. Parker\altaffilmark{7},
A. Ptak\altaffilmark{9},
V. Rana\altaffilmark{1},
D. Stern\altaffilmark{1,10},
N. Webb\altaffilmark{3,4},
W. W. Zhang\altaffilmark{9}
}
\affil{$^{1}$Space Radiation Laboratory, California Institute of Technology, Pasadena,
CA 91125, USA \\
$^{2}$ Department of Astronomy, University of Michigan, 500 Church Street, Ann Arbor, MI 48109-1042, USA \\
$^{3}$ Universite de Toulouse; UPS-OMP; IRAP; Toulouse, France \\
$^{4}$ CNRS; IRAP; 9 Av. colonel Roche, BP 44346, F-31028 Toulouse cedex 4, France \\
$^{5}$ Space Sciences Laboratory, University of California, Berkeley, CA 94720, USA \\
$^{6}$ DTU Space, National Space Institute, Technical University of Denmark, Elektrovej 327, DK-2800 Lyngby, Denmark \\
$^{7}$ Institute of Astronomy, University of Cambridge, Madingley Road, Cambridge CB3 0HA, UK \\
$^{8}$ Columbia Astrophysics Laboratory, Columbia University, New York, NY 10027, USA \\
$^{9}$ NASA Goddard Space Flight Center, Greenbelt, MD 20771, USA \\
$^{10}$ Jet Propulsion Laboratory, California Institute of Technology, Pasadena, CA 91109, USA \\
}

\begin{abstract}
We present results from the coordinated broadband X-ray observations of the
extreme ultraluminous X-ray source \hoix\ performed by \nustar, \xmm\ and
\suzaku\ in late 2012. These observations provide the first high-quality spectra
of \hoix\ above 10\,\kev\ to date, extending the X-ray coverage of this
remarkable source up to $\sim$30\,\kev. Broadband observations were
undertaken at two epochs, between which \hoix\ exhibited both flux and strong
spectral variability, increasing in luminosity from $L_{\rm X} = (1.90 \pm 0.03)
\times 10^{40}$\,\ergps\ to $L_{\rm X} = (3.35 \pm 0.03) \times
10^{40}$\,\ergps. Neither epoch exhibits a spectrum consistent with emission
from the standard low/hard accretion state seen in Galactic black hole binaries,
that would have been expected if \hoix\ harbors a truly massive black hole
accreting at substantially sub-Eddington accretion rates. The \nustar\ data
confirm that the curvature observed previously in the 3--10\,\kev\ bandpass
does represent a true spectral cutoff. During each epoch, the spectrum appears
to be dominated by two optically thick thermal components, likely associated
with an accretion disk. The spectrum also shows some evidence for a
non-thermal tail at the highest energies, which may further support this 
scenario. The available data allow for either of the two thermal components to
dominate the spectral evolution, although both scenarios require highly
non-standard behavior for thermal accretion disk emission.
\end{abstract}


\section{Introduction}

Ultraluminous X-ray Sources (ULXs) are off-nuclear point sources with X-ray
luminosities $L_{\rm X} > 10^{39}$\,\ergps, in excess of the Eddington limit
for the typical 10\,\msun\ stellar-remnant black holes observed in Galactic
black hole binaries (BHBs; \eg \citealt{Orosz03}). Multi-wavelength observations
have largely excluded strong anisotropic emission as a means of substantially
skewing luminosity estimates (\eg \citealt{Berghea10, Moon11}), thus these
high luminosities require either the presence of larger black holes, either
stellar remnant black holes more massive than observed in our own Galaxy (\eg
\cite{Zampieri09}) or perhaps even the long postulated `intermediate mass'
black holes (IMBHs: $10^{2}$ $\lesssim$ \mbh\ $\lesssim$ $10^{5}$\,\msun,
\eg \citealt{Miller04, Strohmayer09a}), or exotic super-Eddington modes of
accretion (\eg \citealt{Poutanen07, Finke07}). The majority of ULXs only radiate
marginally in excess of $10^{39}$\,\ergps\ (\citealt{WaltonULXcat, Swartz11}),
and are likely to simply represent a high luminosity extension of the stellar
mass BHB population (\citealt{Middleton13nat, Liu13nat}). Of particular interest
are the rare population of extreme ULXs with X-ray luminosities $L_{\rm X} >
10^{40}$\,\ergps\ (\eg \citealt{Farrell09, WaltonULXcat, Jonker12, Sutton12,
Walton13culx}). The extreme luminosities displayed by these sources make
them the best candidates for hosting more massive black holes. For recent
reviews on the observational status and the potential nature of ULXs see
\cite{Roberts07rev} and \cite{Feng11rev}.

Previous studies have established that the 0.3--10.0\,\kev\ X-ray spectra of
these extreme ULXs typically show evidence for two separate continuum
components (\eg \citealt{Miller03, Vierdayanti10, Miller13ulx}), one dominating
at softer ($\lesssim$2\,\kev) and the other at harder X-rays ($\gtrsim$2\,\kev), 
potentially analogous to the disk--corona accretion components inferred for
sub-Eddington BHBs (see \citealt{Remillard06rev} for a review). However,
studies focusing on the highest signal-to-noise data have found that the harder
component generally shows evidence of subtle curvature in the
$\sim$3--10\,\kev\ bandpass (\citealt{Stobbart06, Gladstone09, Walton4517}),
which is not observed in the coronal emission of standard sub-Eddington
accretion states. A number of interpretations have since been proposed for this
spectral structure, which can broadly be grouped into those that invoke thermal
processes for the harder component (\eg emission from a hot accretion disk;
\citealt{Gladstone09, Middleton11b, Sutton13uls}), which generally invoke
super-Eddington accretion, and those that invoke non-thermal processes (\eg
a combination of a powerlaw continuum and relativistic reflection from the
inner disk; \citealt{Caball10}), which may still involve intermediate mass black
holes (IMBHs; \mbh\ $\sim 10^{2-5}$\,\msun). As demonstrated in
\cite{Walton4517}, these different model families predict substantially different
spectra above 10\,\kev, in the bandpass only readily accessible with the
focusing hard X-ray telescopes aboard the recently launched Nuclear
Spectroscopic Telescope Array (\nustar; \citealt{NUSTAR}).

\hoix\ is one of the best studied extreme ULXs, which although known to vary
in flux (\eg \citealt{Kong10, Vierdayanti10}) is one of the few sources (within
$\sim$5\,Mpc) to persistently radiate at $L_{\rm X} > 10^{40}$\,\ergps. Early
\xmm\ observations revealed the possible presence of a very cool accretion
disk (\citealt{Miller03}), which may evolve in a fashion similar to the $L \propto
T^{4}$ relation expected for simple blackbody radiation (\citealt{Miller13ulx}),
and indicate the presence of a massive black hole. However, as with other
extreme ULXs for which high quality data are available, \hoix\ shows evidence
for high energy spectral curvature (\eg \citealt{Stobbart06, Gladstone09,
Walton13hoIXfeK}). \hoix\ is also one of the brightest ULXs in the
iron \ka\ bandpass, and sensitive searches have been made for absorption
features that would be indicative of the massive outflows ubiquitously predicted
by simulations of super-Eddington accretion (\eg \citealt{Ohsuga11, Dotan11}).
No features are detected, with limits that require that any undetected features
to be weaker than the iron absorption features resulting from the outflows in a
number of sub-Eddington Galactic BHBs (\citealt{Walton12ulxFeK,
Walton13hoIXfeK}).

Reconciling these various results into a coherent picture regarding the nature of
\hoix\ remains challenging. In order to shed further light onto the nature of the
accretion in this source, we undertook a series of observations with \nustar\ in
order to determine the nature of the high energy emission. These were
coordinated with \suzaku\ and/or \xmm, providing the first high quality
broadband ($\sim$0.3--30.0\,\kev) X-ray spectra of \hoix. The paper is
structured as follows: section \ref{sec_red} describes our data reduction
procedure, and sections \ref{sec_spec} and \ref{sec_var} describe the analysis
performed. We discuss our results in section  \ref{sec_dis} and summarize our
conclusions in section \ref{sec_conc}.

\section{Data Reduction}
\label{sec_red}

\hoix\ was observed by each of the \nustar, \xmm\ and \suzaku\ X-ray
observatories on multiple occasions during 2012. The observations used
in this work are summarized in Table \ref{tab_obs}. Here, we outline our
data reduction for these observations.

\begin{table}
  \caption{Details of the X-ray observations considered in this work, ordered
  chronologically.}
\begin{center}
\begin{tabular}{c c c c}
\hline
\hline
\\[-0.25cm]
Mission & OBSID & Date & Good Exposure\tmark[a] \\
& & & (ks) \\
\\[-0.3cm]
\hline
\hline
\\[-0.1cm]
\multicolumn{4}{c}{\textit{Epoch 1}} \\
\\[-0.2cm]
\suzaku\ & 707019020 & 2012-10-21 & 107 \\
\\[-0.225cm]
\xmm\ & 0693850801 & 2012-10-23 & 6/10 \\
\\[-0.225cm]
\suzaku\ & 707019030 & 2012-10-24 & 107 \\
\\[-0.225cm]
\xmm\ & 0693850901 & 2012-10-25 & 7/13 \\
\\[-0.225cm]
\suzaku\ & 707019040 & 2012-10-26 & 110 \\
\\[-0.225cm]
\nustar\ & 30002033002 & 2012-10-26 & 31 \\
\\[-0.225cm]
\nustar\ & 30002033003 & 2012-10-26 & 88 \\
\\[-0.225cm]
\xmm\ & 0693851001 & 2012-10-27 & 4/13 \\
\\
\multicolumn{4}{c}{\textit{Epoch 2}} \\
\\[-0.2cm]
\nustar\ & 30002033005 & 2012-11-11 & 41 \\
\\[-0.225cm]
\nustar\ & 30002033006 & 2012-11-11 & 35 \\
\\[-0.225cm]
\xmm\ & 0693851701 & 2012-11-12 & 7/9 \\
\\[-0.225cm]
\nustar\ & 30002033008 & 2012-11-14 & 15 \\
\\[-0.225cm]
\xmm\ & 0693851801 & 2012-11-14 & 7/9 \\
\\[-0.225cm]
\nustar\ & 30002033010 & 2012-11-15 & 49 \\
\\[-0.225cm]
\xmm\ & 0693851101 & 2012-11-16 & 3/7 \\
\\[-0.2cm]
\hline
\hline
\\[-0.15cm]
\end{tabular}
\\
$^{a}$ \xmm\ exposures are listed for the \epicpn/MOS detectors, while
\nustar\ exposures quoted are for each of the focal plane modules.
\vspace*{0.3cm}
\label{tab_obs}
\end{center}
\end{table}

\begin{figure*}
\hspace*{-1.0cm}
\epsscale{1.14}
\plotone{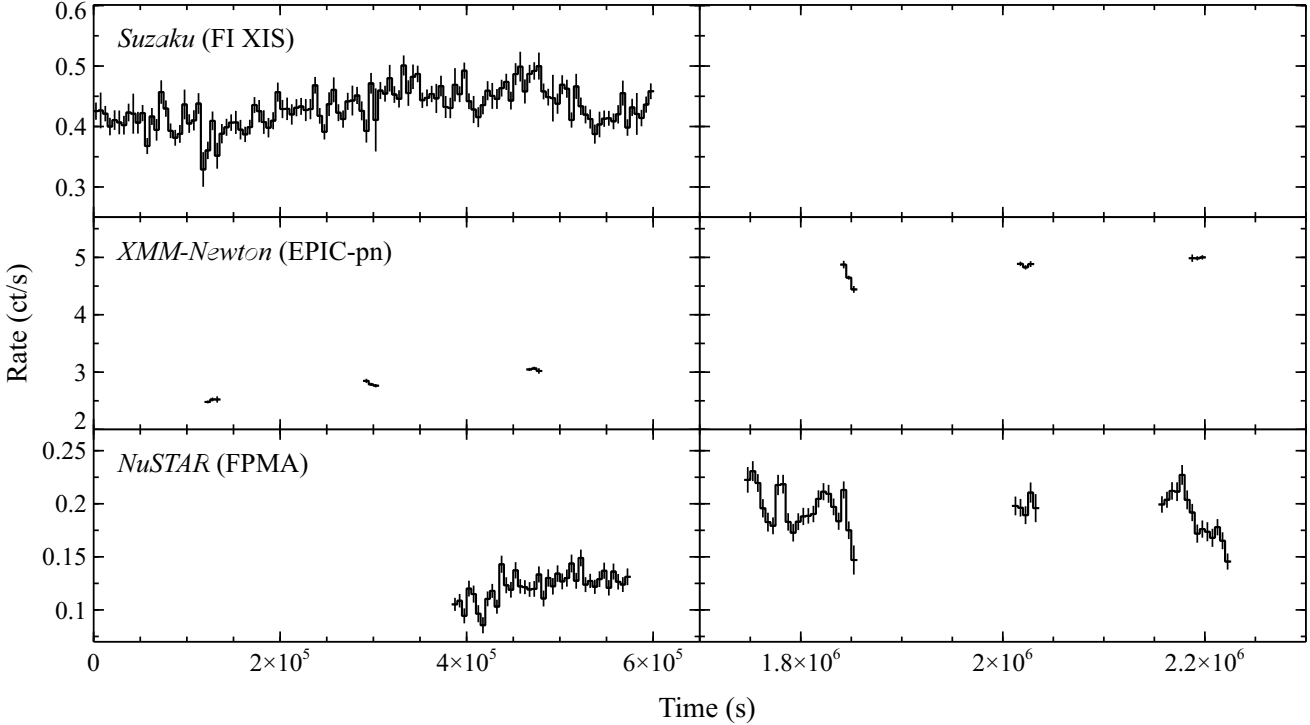}
\caption{
The lightcurves obtained for \hoix\ in 5\,ks bins with \suzaku\ (\textit{top panels}),
\nustar\ (\textit{middle panels}) and \xmm\ (\textit{bottom panels}) during our
observing program in late 2012, indicating the relative coordination of the various
observations utilized in this work.}
\label{fig_lc}
\end{figure*}

\subsection{NuSTAR}

\nustar\ performed two observations of \hoix\ in late 2012 (referred to
throughout this work as epochs 1 and 2), separated by roughly two  weeks (see
Table \ref{tab_obs}). While the first observation (OBSIDs 30002033002 and
30002033003) was taken continuously, the second (OBSIDs 30002033005, 
30002033006, 30002033008 and 30002033010) was split into three
contemporaneous segments in order to maximise the overlap with our \xmm\
observations (see below). We reduced the \nustar\ data using the standard
pipeline, part of the \nustar\ Data Analysis Software v1.3.0 (\nustardas;
included in the standard \heasoft\ distribution as of version 14), and we used
instrumental responses from \nustar\ caldb v20131007 throughout. We
cleaned the unfiltered event files with the standard depth correction, which
significantly reduces the internal background at high energies, and
removed periods of earth-occultation and passages through the South Atlantic
Anomaly, which result in a typical observing efficiency of $\sim$50\%. Source
products were obtained from circular regions (radius $\sim$70$''$), and the
background was estimated from a larger, blank area of the same detector free
of contaminating point sources. Spectra and lightcurves were extracted from
the cleaned event files using \xselect\ for both focal plane modules (FPMA and
FPMB). Finally, the spectra were grouped such that each spectral bin contains at
least 50 counts. These \nustar\ observations provide good spectra for \hoix\
up to $\sim$30--35\,\kev\ from each of the focal plane modules, FPMA and
FPMB, for each epoch. In this work, we fit the spectra from FPMA and FPMB
jointly, without combining them.

\subsection{Suzaku}

\subsubsection{XIS Detectors}

The first \nustar\ observation of \hoix\ was performed simultaneously with a
portion of our recent long integration (\citealt{Walton13hoIXfeK}) with the
\suzaku\ observatory (\citealt{SUZAKU}). Figure \ref{fig_lc} highlights the
relative coordination of all the observations considered in this work. The data
reduction procedure for the XIS detectors (\citealt{SUZAKU_XIS}) for the full
\suzaku\ dataset, including observations taken in April 2012, has already
been described in \cite{Walton13hoIXfeK}, following the procedure
recommended in the \suzaku\ data reduction
guide.\footnote{http://heasarc.gsfc.nasa.gov/docs/suzaku/analysis/} Here
we only consider the \suzaku\ data taken either simultaneously or
contemporaneously with \nustar\ in October, and we re-reduce the spectra
following the same procedure with the latest XIS calibration files (released
September 2013), which substantially improve the agreement between the
front- and back-illuminated XIS detectors at low energies. In this work, we
model the XIS data over the 0.7--10.0\,\kev\ energy range (unless stated
otherwise), excluding the 1.6--2.1\,\kev\ band throughout owing to
remaining calibration uncertainties associated with the instrumental silicon
K edge. We also rebin the XIS spectra to have a minimum of 50 counts per
energy bin.

\subsubsection{HXD PIN}

Owing to the combination of the systematic uncertainty in the background
model for the \suzaku\ PIN detector (\citealt{SUZAKU_HXD}) -- equivalent
to $\gtrsim$25\% of the `source' flux for the weak detection of the
Holmberg\,IX field (see \eg discussion in \citealt{Walton13spin}) -- and the
source confusion resulting from its lack of imaging capability -- the variable
nucleus of M\,81 (\eg \citealt{Markoff08, JMiller10}) also falls in the PIN
field-of-view -- the data obtained with this detector unfortunately cannot
be used to constrain the high energy ($E > 10$\,\kev) properties of \hoix\
(\citealt{Walton13hoIXfeK}). Therefore, we do not consider the PIN data here,
and focus instead on the high energy data provided by \nustar.

\subsection{XMM-Newton}

For our \xmm\ observations, data reduction was carried out with the \xmm\
Science Analysis System (\sas v13.5.0) largely according to the standard
prescription provided in the online guide\footnote{http://xmm.esac.esa.int/}.
The observation data files were processed using \epchain\ and \emchain\ to
produce calibrated event lists for the \epicpn\ (\citealt{XMM_PN}) and \epicmos\
(\citealt{XMM_MOS}) detectors respectively. Source products were extracted from
circular regions of radius $\sim$40$''$ for \epicpn, and of radius $\sim$50$''$
for the \epicmos, selected to avoid chip gaps (where relevant). In each case the
background was estimated from a larger area of the same CCD free from
contaminating point sources. Lightcurves and spectra were generated with
\xmmselect, excluding periods of high background flares and selecting only 
single and double events for \epicpn, and single to quadruple events for
\epicmos\ during the first epoch. For the second epoch, the source was bright
enough for the \epicmos\ data to be affected by mild pileup, and so for these
three observations we considered only single grade events in order to mitigate
against these effects. The redistribution matrices and auxiliary response files
were generated with \rmfgen\ and \arfgen, while lightcurves were corrected for
the background count rate using \epiclccorr. After performing the data reduction
separately for each of the MOS detectors, and confirming their consistency, these
spectra were combined using the \ftool\ \addascaspec. Finally, spectra were
re-binned using the SAS task \specgroup\ to have a minimum of 50 counts in
each energy bin. The \xmm\ data are modeled over the full 0.3--10.0\,\kev\
bandpass (unless stated otherwise).

\begin{figure*}
\hspace*{-0.7cm}
\epsscale{1.1}
\plotone{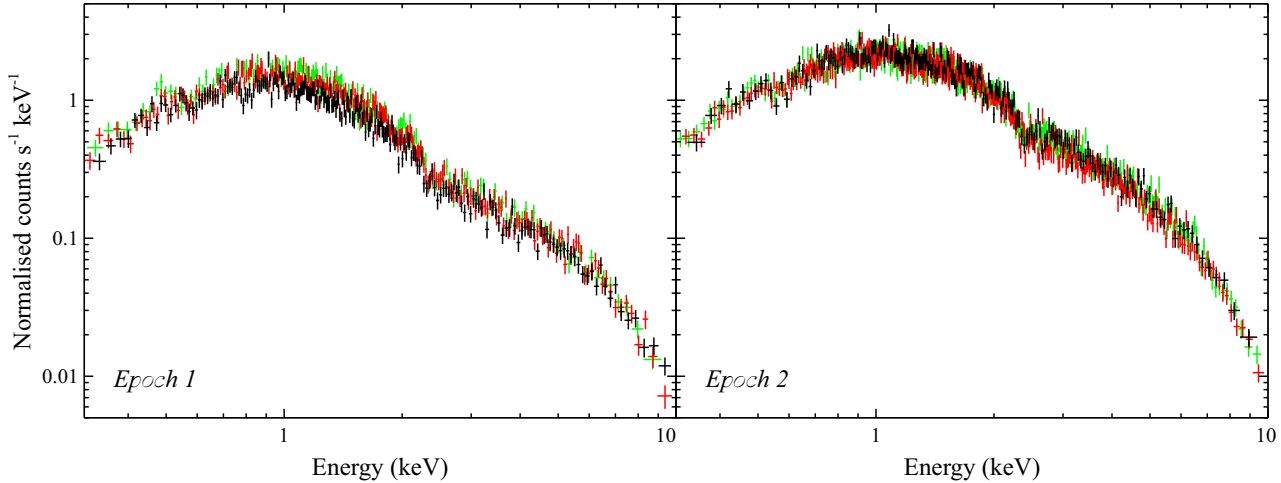}
\caption{
\epicpn\ spectra for the two sets of three observations of \hoix, the first taken
in October (\textit{left panel}; associated with \nustar\ observation 1) and the
second in November (\textit{right panel}; associated with \nustar\ observation 2).
The black, red and green spectra show the first, second and third observations
from each set of three. Although there is some mild flux variability displayed by
the observations in each set, it is clear there is very little spectral variability.}
\label{fig_xmmspec}
\end{figure*}

\section{Spectral Analysis}
\label{sec_spec}

During these observations, \hoix\ displayed some flux variability, both between
epochs and within a single epoch (Figure \ref{fig_lc}). We therefore test to see
whether this flux variability is accompanied by strong spectral variability in order
to determine how to best extract spectra. In Figure \ref{fig_xmmspec} we show
the \xmm\ spectra obtained from each of the three observations associated with
each epoch. Despite the mild flux variability, the spectra obtained within each
epoch show good consistency. However, we do see differences in the broadband
spectra when comparing epoch 1 and epoch 2. Given the lack of short-term
spectral variability, we do not require strict simultaneity between \nustar,
\suzaku\ and/or \xmm. This allows us to maintain the highest S/N possible in
the \nustar\ data, and we combine the available data from each of the missions
into average spectra for each epoch, which we analyze simultaneously. However,
we keep our analysis of epochs 1 and 2 separate.

We investigate the broadband spectral properties of \hoix\ utilizing a similar
suite of models to those applied in recent broadband spectral studies of other
ULXs (\citealt{Walton13culx, Bachetti13, Rana14}, \textit{submitted}).
Throughout this work, spectral modelling is performed with XSPEC v12.8.0
(\citealt{XSPEC}), and quoted uncertainties on spectral parameters are the 90
per cent confidence limits for a single parameter of interest, unless stated
otherwise. Neutral absorption is treated with \tbnew\footnote{\tbnewlink}, the
latest version of the \tbabs\ absorption code (\citealt{tbabs}), with the
appropriate solar abundances, and the absorption cross-sections of
\cite{Verner96}. All models include Galactic absorption with a column of
$N_{\rm H; Gal} = 5.54 \times 10^{20}$\,cm$^{-2}$ (\citealt{NH}), in addition
to an intrinsic neutral absorber at the redshift of Holmberg\,IX ($z$ =
0.000153) with a column that is free to vary (unless stated otherwise).

\subsection{Epoch 1}

\subsubsection{Cross-Calibration}
\label{sec_crosscal}

\begin{figure}
\hspace*{-0.75cm}
\epsscale{1.16}
\plotone{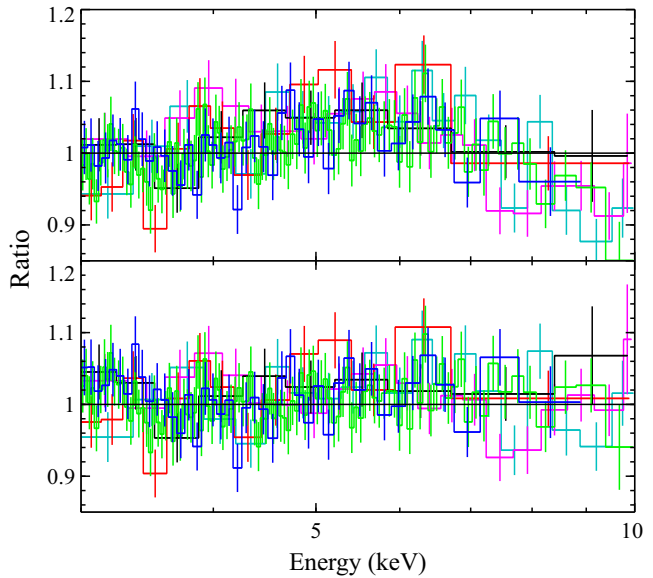}
\caption{
Data/model ratios for the \xmm\ (black: \epicpn, red: \epicmos), \suzaku\
(green: FI XIS, blue: BI XIS) and \nustar\ (magenta: FPMA, cyan: FPMB) datasets
obtained during epoch 1, modelled with both a powerlaw continuum (\textit{top
panel}) and a bremsstrahlung continuum (\textit{bottom panel}). The data from
the various missions display consistent curvature across their common energy
range (3--10\,\kev). The data have been rebinned for visual clarity.}
\label{fig_crosscal}
\end{figure}

We focus initially on the data obtained during the first epoch, and first
investigate the agreement between the three missions over their common
3--10\,\kev\ bandpass, taking this epoch to be representative. We first apply
an unabsorbed powerlaw model to this energy range for each of the three
observatories, allowing multiplicative constants to float between the spectra
obtained from the various detectors (we choose the constant for \nustar\ FPMA
to be unity). Although this model formally provides a good fit, with \rchi\ =
1984/1928 and $\Gamma_{\rm 3-10~keV} = 1.81^{+0.01}_{-0.02}$,
systematic curvature in the residuals can be seen across this bandpass (see
Figure \ref{fig_crosscal}, \textit{top panel}). Parameterising the data with a
curved continuum instead, using a simple unabsorbed bremsstrahlung model,
provides an excellent fit (Figure \ref{fig_crosscal}, \textit{bottom panel}), with
\rchi\ = 1870/1928 and $T = 11.0 \pm 0.3$\,\kev, an improvement of
$\Delta\chi^{2}$ = 114 (for no additional degrees of freedom) over the
powerlaw continuum. Allowing the temperatures to vary independently for
each of the different missions does not significantly improve the fit (\rchi\ =
1861/1926), and the temperatures obtained all agree within 2$\sigma$ or
better: $T_{\rm XMM} = 12.7^{+1.8}_{-1.3}$\,\kev, $T_{\rm Suzaku} = 11.1
\pm 0.4$\,\kev\ and $T_{\rm NuSTAR} = 10.4 \pm 0.5$\,\kev. Therefore, we
conclude the spectra obtained with \xmm, \suzaku\ and \nustar\ show good
agreement (see also \citealt{Walton13culx}, Madsen et al. 2014; \textit{in
preparation}). The fluxes obtained for each of the different detectors all agree
with that of \nustar\ FPMA to better than $\sim$10\%.

Furthermore, the preference for the bremsstrahlung continuum over the
powerlaw confirms the presence of curvature in the observed 3--10\,\kev\
spectrum, similar to that inferred from earlier \xmm\ data (\citealt{Stobbart06,
Gladstone09}), and from the full \suzaku\ dataset (including additional data
obtained earlier in 2012, \citealt{Walton13hoIXfeK}). The absorption column
towards \hoix\ is $N_{\rm H} \lesssim 2 \times 10^{21}$\,cm$^{-2}$ (\eg
\citealt{Miller13ulx}), which is not sufficient to significantly influence the
spectrum in the 3--10\,\kev\ bandpass. Thus, the observed curvature must be
intrinsic to the 3--10\,\kev\ continuum.

\begin{figure}
\hspace*{-0.75cm}
\epsscale{1.16}
\plotone{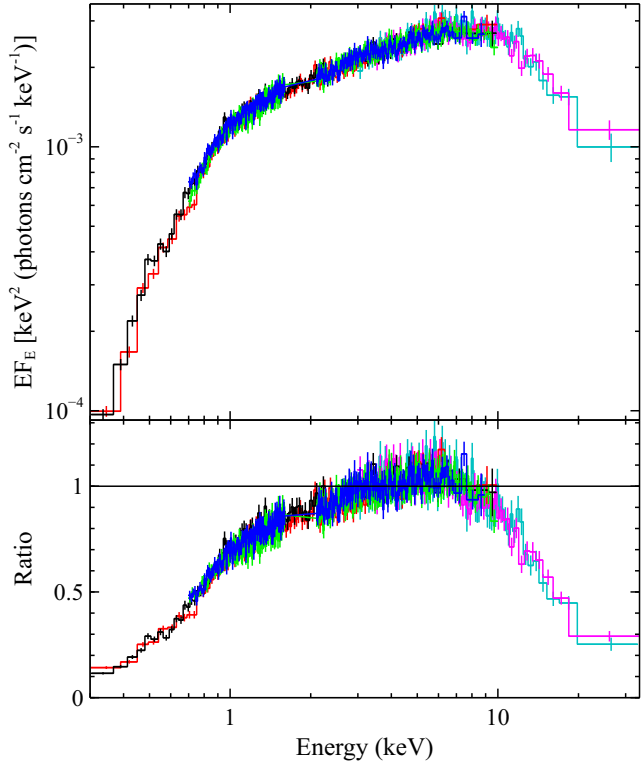}
\caption{
\textit{Top panel:} The broadband X-ray spectrum displayed by \hoix\ during
epoch 1. All the data have been unfolded through the same model, which simply
consists of a count spectrum that is constant with energy. \textit{Bottom panel:}
data/model ratio to the powerlaw continuum applied to the 3--10\,\kev\ energy
range, then extended out to the full broadband spectrum. Both panels clearly
demonstrate that the curvature observed in the 3--10\,\kev\ bandpass is
associated with a true spectral cutoff. The color scheme is the same as for Figure
\ref{fig_crosscal}, and the data have again been rebinned for clarity.}
\label{fig_obs1}
\end{figure}

\begin{figure}
\hspace*{-0.75cm}
\epsscale{1.16}
\plotone{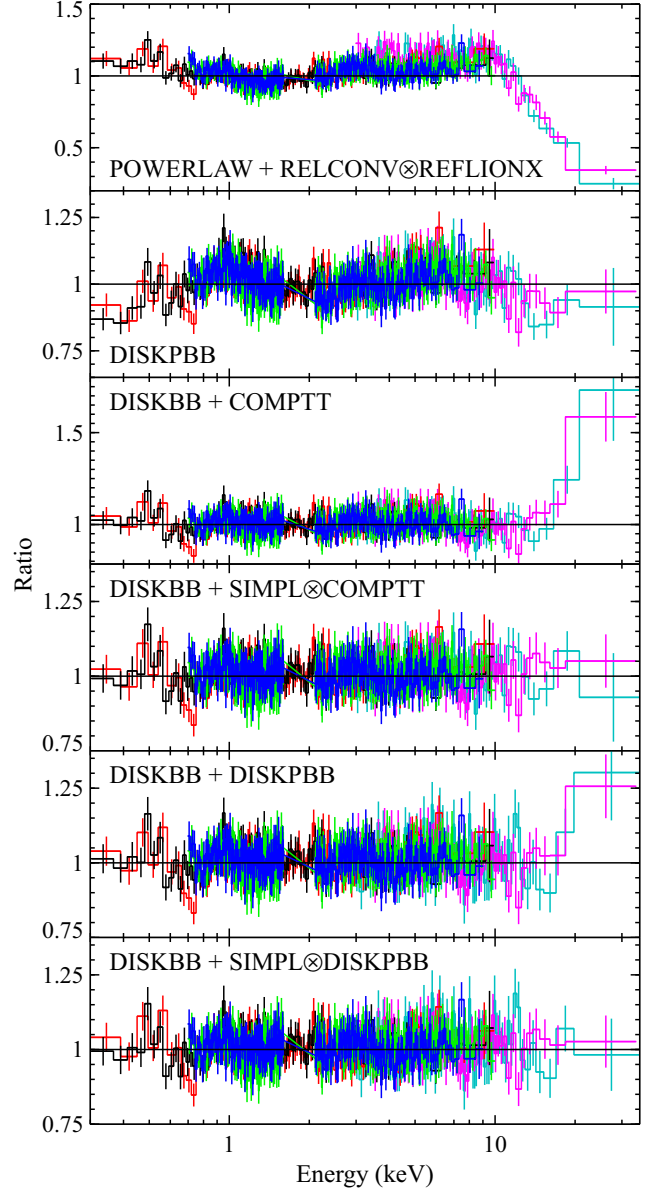}
\caption{
Data/model ratios for a variety of the continuum models applied to the full
broadband X-ray spectrum of \hoix\ obtained during epoch 1 (see section
\ref{sec_epoch1}). The color scheme is the same as for Figures \ref{fig_crosscal}
and \ref{fig_obs1}, and the data have again been rebinned for visual clarity.}
\label{fig_obs1_ratio}
\end{figure}

\subsubsection{Broadband Continuum Modelling}
\label{sec_epoch1}

We now analyse the full broadband \xmm\ + \suzaku\ + \nustar\ spectrum
obtained from the first epoch, applying a suite of continuum models in order to
examine the nature of the broadband emission from \hoix. Figure  \ref{fig_obs1}
(\textit{top panel}) shows the spectra from epoch 1 unfolded through a model
that simply consists of a constant. We also show in Figure \ref{fig_obs1}
(\textit{bottom panel}) the data/model ratio to the powerlaw continuum initially
applied to the 3--10\,\kev\ bandpass (as outlined in section \ref{sec_crosscal}),
then extrapolated across the full 0.3--30.0\,\kev\ spectrum considered here.

\begin{table*}
  \caption{Best fit parameters obtained for the variety of continuum models
  applied to the broadband data available for \hoix.}
\begin{center}
\begin{tabular}{c c c c c c c c}
\hline
\hline
\\[-0.2cm]
Model & $N_{\rm H; int}$ & $kT_{\rm in,1}$ & $kT_{\rm e}$ or $kT_{\rm in,2}$\tmark[a] & $p$ or $\tau$ & $\Gamma$ & $f_{\rm scat}$ & $\chi^{2}$/DoF \\
\\[-0.25cm]
& ($10^{21}$ cm$^{-2}$) & (\kev) & (\kev) & & & (\%) & \\
\\[-0.25cm]
\hline
\hline
\\[-0.1cm]
\multicolumn{8}{c}{\textit{Epoch 1}} \\
\\[-0.15cm]
\diskpbb & $1.27^{+0.03}_{-0.04}$ & $4.87^{+0.13}_{-0.12}$ & - & $0.542 \pm 0.001$ & - & - & 3913/3522 \\
\\[-0.2cm]
\diskbb+\comptt & $1.5^{+0.2}_{-0.3}$ & $0.15 \pm 0.01$ & $3.1 \pm 0.1$ & $6.3 \pm 0.1$ & - & - & 3651/3520 \\
\\[-0.2cm]
\diskbb+\simpl${\otimes}$\comptt & $1.4^{+0.1}_{-0.2}$ & $0.23 \pm 0.04$ & $2.4^{+0.3}_{-0.4}$ & $7.3^{+0.4}_{-0.5}$ & $>2.4$ & $>16$ & 3581/3518 \\
\\[-0.2cm]
\diskbb+\diskpbb\ & $1.7 \pm 0.1$ & $0.26 \pm 0.02$ & $4.2 \pm 0.1$ & $0.560^{+0.004}_{-0.003}$ & - & - & 3606/3520 \\
\\[-0.2cm]
\diskbb+\simpl$\otimes$\diskpbb\ & $1.6 \pm 0.1$ & $0.27^{+0.01}_{-0.02}$ & $3.8 \pm 0.2$ & $0.564 \pm 0.004$ & $1.63^{+0.09}_{-0.08}$ & $3 \pm 1$ & 3584/3518 \\
\\
\multicolumn{8}{c}{\textit{Epoch 2}} \\
\\[-0.15cm]
\diskpbb & $1.67 \pm 0.06$ & $2.56 \pm 0.03$ & - & $0.587 \pm 0.004$ & - & - & 2166/1584 \\
\\[-0.2cm]
\diskbb+\comptt\ & $0.59 \pm 0.04$ & $1.17 \pm 0.06$ & $2.7 \pm 0.1$ & $7.1^{+0.7}_{-0.5}$ & - & - & 1949/1582 \\
\\[-0.2cm]
\diskbb+\simpl${\otimes}$\comptt\ & $1.3^{+0.2}_{-0.1}$ & $0.3 \pm 0.1$ & $1.11^{+0.06}_{-0.04}$ & $11.3 \pm 0.6$ & $3.6 \pm 0.1$ & $>81$ & 1695/1580 \\
\\[-0.2cm]
\diskbb+\diskpbb\ & $1.7 \pm 0.1$ & $1.79 \pm 0.04$ & $5.4^{+0.5}_{-0.4}$ & $0.514 \pm 0.007$ & - & - & 1693/1582 \\
\\[-0.2cm]
\diskbb+\simpl$\otimes$\diskpbb\ & $1.5^{+0.3}_{-0.1}$ & $1.63^{+0.07}_{-0.06}$ & $3.5^{+0.6}_{-0.3}$ & $0.55^{+0.01}_{-0.02}$ & $<1.65$ & $2.3^{+0.9}_{-0.6}$ & 1686/1580 \\
\\[-0.2cm]
\hline
\hline
\end{tabular}
\\[0.125cm]
$^{a}$ This column gives the temperature of the hotter continuum component,
where relevant, i.e. $kT_{\rm e}$ for models including a \comptt\ component,
or the inner temperature $kT_{\rm in}$ of the \diskpbb\ component for models
invoking both \diskbb\ and \diskpbb.
\label{tab_param}
\end{center}
\end{table*}

The \nustar\ data clearly show that the curvature displayed in the 3--10\,\kev\
bandpass extends to higher energy, and genuinely represents a spectral cutoff,
similar to NGC\,1313 X-1 (\citealt{Bachetti13}) and IC\,342 X-1
(\citealt{Rana14}, \textit{submitted}). This is contrary to the basic expectation
for the interpretation in which the 3--10\,\kev\ curvature is produced by
relativistic disk reflection with the accretion disk illuminated by a standard
sub-Eddington powerlaw-like corona (\citealt{Caball10}), as the Compton
hump  should cause the spectrum to turn back up at higher energies
($\gtrsim$10\,\kev; \citealt{Walton4517}). This model provides a good fit to
the data below 10\,\kev\ (\rchi\ = 3419/3360), utilizing the \reflionx\ code
(\citealt{reflion}) for the reflected emission and the \relconv\ kernel
(\citealt{relconv}) to account for the relativistic effects inherent to the inner
accretion flow around a black hole. However, when applying this model to the
full 0.3--30.0\,\kev\ bandpass we see that when the curvature is modeled
as being due to iron emission, the data are significantly overpredicted at the
highest energies (Figure \ref{fig_obs1_ratio}), and the resulting broadband fit is
poor (\rchi\ = 6016/3516). We therefore proceed by considering models that
invoke a thermal origin for the high energy curvature.

Simple multi-color blackbody accretion disk models (\eg \diskbb:
\citealt{DISKBB}, \diskpn: \citealt{DISKPN}), assuming a geometrically thin,
optically thick disk as per \cite{Shakura73}, also fail to fit the broadband data
(\diskbb: \rchi\ = 22682/3523; \diskpn: \rchi\ = 20303/3522). Although the
high energy data do show a cutoff similar to a thermal spectrum, the overall
profile is too broad to be explained by such simple models. However, at very
high- or super-Eddington rates the accretion disk may differ substantially from
the simple \cite{Shakura73} thin disc profile owing to the increased effects of
radiation pressure and advection (\citealt{Abram88}). This can result in shallower
radial temperature profiles and the appearence of a broader, less-peaked
emission profile from the disk. Allowing the radial temperature profile of the disk
($p$) to vary as an additional free parameter using the \diskpbb\ model
(\citealt{diskpbb}) does substantially improve the fit (\rchi\ = 3913/3522), and
the inferred temperature profile is shallower than expected for a thin disk (i.e. $p
<$ 3/4); the results are presented in Table \ref{tab_param}. However, significant
structure remains in the residuals (see Figure \ref{fig_obs1_ratio}), and the
spectrum does appear to require at least two separate continuum components.

We therefore model the broadband spectrum from epoch 1 with the
\diskbb+\comptt\ combination frequently used to parameterize the spectra
from bright ULXs below 10\,\kev\ (\eg \citealt{Gladstone09, Middleton11b,
Walton4517}). This model is used to represent a disk--corona accretion
geometry, in which hot coronal electrons Compton up-scatter seed photons
from the (cooler) accretion disk into a second hard emission component,
modelled here with \comptt\ (\citealt{comptt}). For simplicity, we require the
temperature of the seed photons to be that of the accretion disk component.
Although this may not be appropriate if this emission genuinely arises
from an optically-thick corona (as frequently inferred for ULXs with this model),
which may obscure our view of the true inner disk (\eg \citealt{Gladstone09,
Pintore14}), it is sufficient for our purposes as relaxing this assumption does
not improve the fit, nor change the parameter values obtained. The \comptt\
model allows both the temperature and the optical depth of the Comptonizing
region to be fit as free parameters. This does offer a significant improvement
over the simpler \diskpbb\ model, with \rchi\ = 3651/3520, and the results
obtained are fairly similar to previous applications of this model to ULX spectra
(see Table \ref{tab_param}), \ie the electron plasma is inferred to be fairly cool
($T_{\rm e} \sim 3$\,\kev) and optically thick ($\tau \sim 6$) owing to the
observed spectral curvature in the $\sim$3--10\,\kev\ bandpass, and the
seed photon/disk temperature is very cool ($T_{\rm in} \sim 0.15$\,\kev).
However, this model leaves a significant excess at the highest energies probed
by \nustar\ (see Figure \ref{fig_obs1_ratio}), requiring a further high energy
continuum component.

Although the \diskbb+\comptt\ model formally describes a disk--corona
scenario, when applied to ULX spectra the \comptt\ component has frequently
been interpreted as representing emission from a distorted high-Eddington
accretion disk (\eg \citealt{Middleton11b}) rather than standard coronal
emission, owing to the large optical depth typically inferred, and the unusually
low electron temperatures compared to those observed from standard
sub-Eddington coronae (\eg \citealt{Gierlinski99, Miller13grs, Tomsick14,
Natalucci14, Brenneman14}). In this scenario, it is therefore plausible that the
high energy excess observed represents the true emission from the
Comptonizing corona. The addition of a powerlaw tail to the \comptt\
component included in this model, using the \simpl\ convolution model
(\citealt{simpl}; limiting the photon index to $1.5 \leq \Gamma \leq 4$, broadly
encompassing the range observed from Galactic binaries), provides a further
significant improvement of $\Delta\chi^{2} = 70$ (for two additional free
parameters) to the fit, and resolves the high energy excess (see Figure
\ref{fig_obs1_ratio}). Even if poorly constrained owing to a strong degeneracy
with the scattered fraction $f_{\rm scat}$ (which serves as the \simpl\
`normalisation'; see Table \ref{tab_param}), the photon index is very steep,
$\Gamma > 2.4$. We stress that even though this model still utilizes
\comptt\ to model the $\sim$3--10\,\kev\ emission, the necessity for
a second Comptonizing region at even higher energies strongly suggests that
this emission does arise from a hot, super-Eddington disk, which the \comptt\
model merely has the flexibility to mimic, rather than physically being
associated with an optically-thick corona.

\begin{figure}
\hspace*{-0.75cm}
\epsscale{1.16}
\plotone{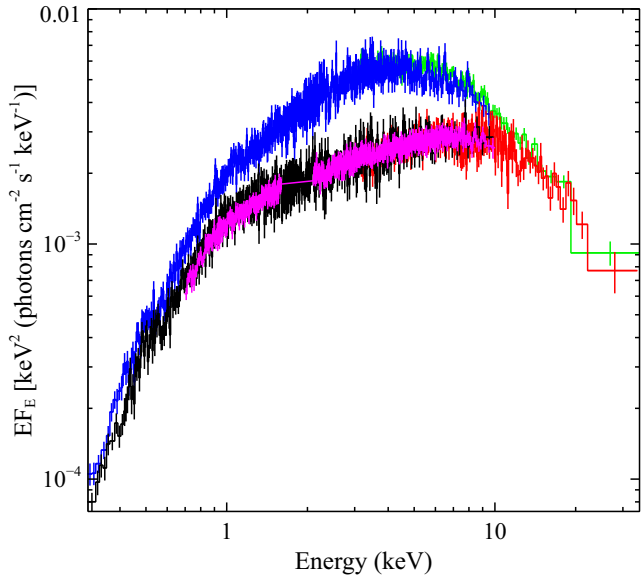}
\caption{
The spectral evolution displayed by \hoix\ between the two epochs. The \xmm\
(\epicpn), \suzaku\ (FI XIS) and \nustar\ (FPMA) data from epoch 1 are shown in
black, magenta and red, respectively, while the \xmm\ and \nustar\ data from
epoch 2 are shown in blue and green. All the data have been unfolded
through the same model, which again consists of a constant.}
\label{fig_eeuf}
\end{figure}

Finally, if the high energy emission is dominated by a hot accretion disk,
describing this emission with a \comptt\ component with the seed photon
temperature linked to the soft thermal component might not be correct. We
therefore also test other models composed of two thermal components in
order to test whether the hard excess observed with the \diskbb+\comptt\
combination is simply a consequence of that particular model. A model
consisting of two \diskbb\ components provides a poor fit to the data (\rchi\
= 4228/3521), leaving a strong excess in the \nustar\ data above 10\,\kev\
and substantial residual structure at lower energies. Replacing the second
(hotter) \diskbb\ with a \diskpbb\ component, with the radial temperature
profile free to vary again provides a significant improvement (\rchi\ =
3606/3520). However, there is still evidence for an excess remaining at the
highest energies (see Figure \ref{fig_obs1_ratio}); the addition of \simpl\ here
gives an improvement of $\Delta\chi^{2} = 22$ (for two additional free
parameters). Although the best fit photon index of the powerlaw tail
is rather hard in this case, we note that there is a local minimum of similar
statistical quality (\rchi\ = 3588/3518) with a steep photon index ($\Gamma
> 3.1$). Furthermore, even in the model with the hard photon index, the hard
powerlaw continuum only begins to dominate the model outside of the
bandpass probed. Within the \nustar\ bandpass, the effect is still to append a
steep tail onto the \diskpbb\ component, similar to the case with the
steep photon index, as the model begins to curve up into the hard portion of
this continuum. It seems that the marginal preference for a hard photon index
arises from the subtle difference this model has on the curvature at lower
energies, rather than the slope of the \textit{observed} continuum above
$\sim$15\,\kev, which Figure \ref{fig_obs1} shows is steep.

\begin{figure}
\hspace*{-0.75cm}
\epsscale{1.17}
\plotone{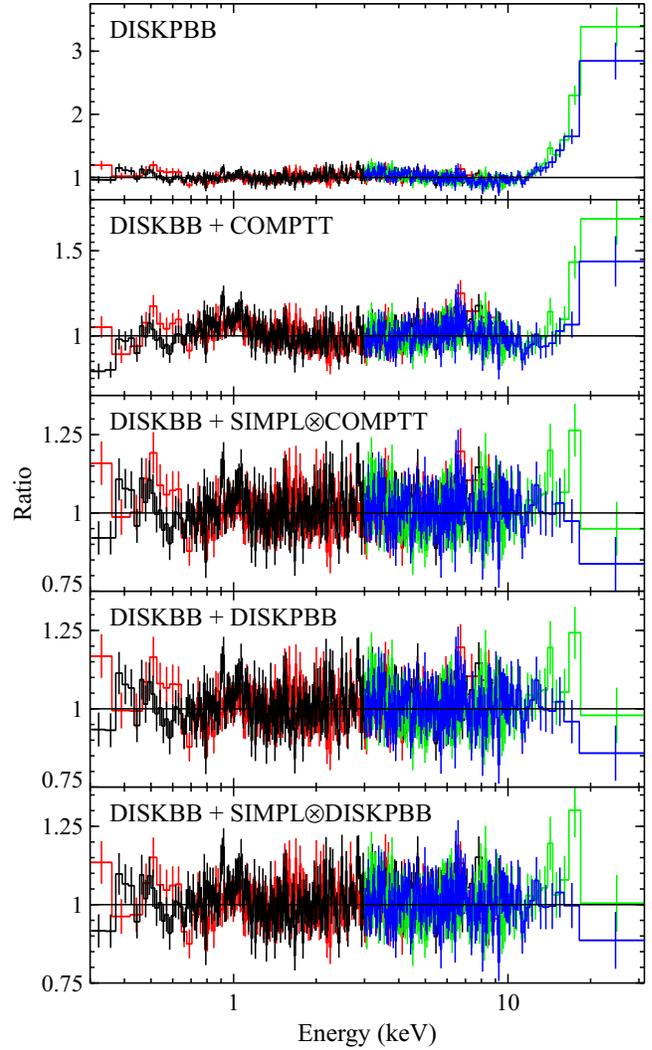}
\caption{
Data/model ratios for a variety of the continuum models applied to the full
broadband X-ray spectrum of \hoix\ obtained during epoch 2 (see section
\ref{sec_epoch2}). \xmm\ \epicpn\ and \epicmos\ data are shown in black
and red, respectively, and \nustar\ FPMA and FPMB in green and blue; the
data have been rebinned for visual clarity.}
\label{fig_obs2_ratio}
\end{figure}

\begin{figure*}
\hspace*{-0.9cm}
\epsscale{1.14}
\plotone{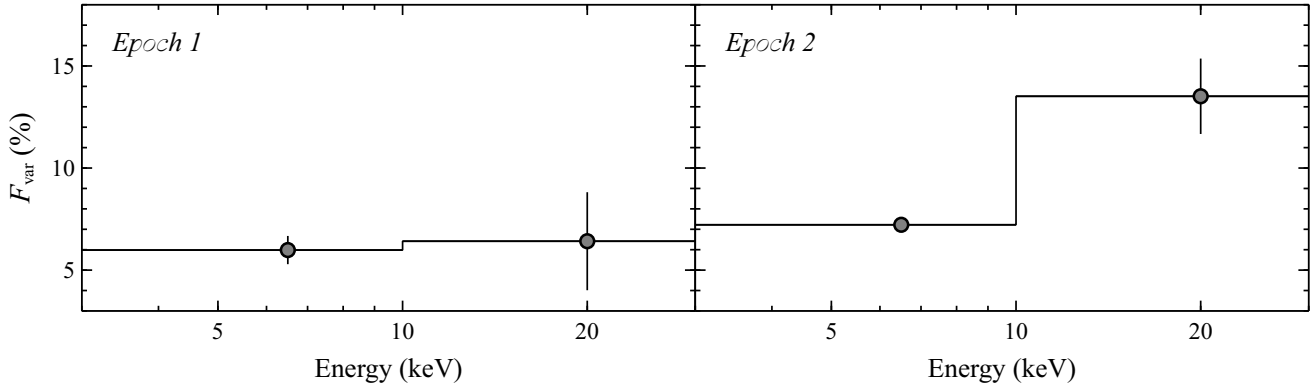}
\caption{
The fractional excess variability (\fvar; covering frequencies $1.4 \times
10^{-5} - 10^{-3}$\,Hz, \ie 5--70\,\ks\ timescales) above and below
10\,\kev\ observed from \hoix\ with \nustar\ during the two epochs (see
section \ref{sec_var}). There is tentative evidence for enhanced variability
above 10\,\kev\ in the second epoch.}
\label{fig_fvar}
\end{figure*}

\subsection{Epoch 2}
\label{sec_epoch2}

During the second epoch, \hoix\ was substantially brighter than during the
first (see Table \ref{tab_flux}). Although there is no evidence for substantial
spectral variability during either of the two epochs, the spectra from the two
epochs do exhibit marked differences, as shown in Figure \ref{fig_eeuf}. The
strongest variability can clearly be seen at $\sim$3\,\kev, and the
3--10\,\kev\ spectrum now appears to be more peaked than during epoch 1.
In contrast, however, there is remarkably little variability in the spectrum at
the highest ($\gtrsim$15\,\kev) and the lowest ($\lesssim$1\,\kev) energies.
One of the thermal components required in epoch 1 appears to have evolved
substantially both in terms of its temperature and its flux, while the other
appears to have remained relatively stable.

We therefore apply the thermal models considered for epoch 1 to these data,
in order to investigate which of these can adequately reproduce the epoch 2
spectrum. Even though the spectrum now appears to be primarily dominated
by one of the thermal components, simple accretion disk models still fail to
adequately reproduce the data; an absorbed \diskbb\ model gives a very poor
fit, with \rchi\ = 4727/1585. Allowing the radial temperature profile to vary
again substantially improves the fit (\rchi\ = 2166/1584), but the data still
show a strong excess over the model at the highest energies, and at least two
continuum components still appear to be required for this epoch (see Figure
\ref{fig_obs2_ratio}).

We therefore fit the \diskbb+\comptt\ model considered previously. While the
fit is improved over the \diskpbb\ model, it is still relatively poor (\rchi\ =
1949/1582), and the high energy excess persists, even though in this model it
is the \diskbb\ component that shifts up in temperature to try and account for
the peak of the emission (at $\sim$4\,\kev; Figure \ref{fig_eeuf}), while the
best fit temperature and the optical depth for the \comptt\ component are
similar to the first epoch (see Table \ref{tab_param}). Allowing again for a
further high energy powerlaw tail to the \comptt\ component with \simpl\
substantially improves the fit (\rchi\ = 1695/1580), and accounts for  the high
energy excess. However, the best fit results are very different to those obtained
with the simpler \diskbb+\comptt\ model. Now it is the \diskbb\ component
that appears to have remained fairly constant, while the \comptt\ component
has decreased its temperature and increased its flux to account for the
$\sim$4\,\kev\ peak. Other aspects have remained similar to the first epoch
though, and the \comptt\ component is again inferred to be optically thick
($\tau \sim 11$).

Finally, we also reconsider the \diskbb+\diskpbb\ model applied previously
to the first epoch. This model provides a similar quality fit to the
\diskbb+\comptt+\simpl\ combination (\rchi\ = 1693/1582), but here it is
the \diskpbb\ component which primarily accounts for the high energy
emission, while the temperature of the \diskbb\ component has increased
substantially to account for the $\sim$4\,\kev\ spectral peak (again, see Table
\ref{tab_param}). In contrast to the first epoch, adding a high energy powerlaw
tail to the \diskpbb\ component only provides a marginal improvement to the
fit (\rchi\ = 1686/1580). Furthermore, we note that as with the first
epoch, even though the best fit photon index is rather hard, this portion of the
continuum again only dominates outside of the observed bandpass, and the
observed continuum at the highest energies probed is still actually very steep,
as is clear from Figure \ref{fig_eeuf}.

\section{Short Term Variability}
\label{sec_var}

We also investigate the short-term variability behavior during these epochs. In
order to crudely assess how the variability may evolve with energy, we produce
lightcurves over the energy ranges 3--10 and 10--30\,\kev\ for each of the
\nustar\ observations, and estimate the fractional excess variability (\fvar;
\citealt{Edelson02, Vaughan03}) for each energy band for the two epochs. In
order to ensure the same timescales are probed during each epoch, we split
the lightcurves into $\sim$70\,ks segments, calculate \fvar\ for each and
compute the weighted average. We use 70\,ks segments as this is roughly the
duration of the last part of the second observation, and it divides the first
observation neatly into three segments. As the central part of observation 2
does not span a sufficient duration, these data are not considered here, and we
only utilize the first 70\,ks of the first part. In order to ensure the variability
above 10\,\kev\ is not dominated by Poisson noise, we again use time bins of
5\,ks. The results are shown in Figure \ref{fig_fvar}. During the first epoch,
mild  variability is observed, which seems to be constant with energy. However,
during the second epoch, while the variability below 10\,\kev\ is broadly
consistent with the level displayed during the first, we see some evidence for
enhanced variability above 10\,\kev, both in comparison to the variability
below 10\,\kev\ during this epoch, and the variability observed during the first
epoch.

\section{Discussion}
\label{sec_dis}

We have presented an analysis of the broadband X-ray spectrum of the extreme
ULX \hoix, observed twice during 2012, by \nustar, \xmm\ and \suzaku.
\nustar\ has provided the first high quality hard X-ray ($E > 10$\,\kev) spectra
of this remarkable source to date. During both epochs the hard X-ray emission
is weak in comparison to the soft X-ray ($E<10$\,\kev) emission, as
demonstrated by the fluxes presented in Table \ref{tab_flux}; the flux above
10\,\kev\ represents at most $\sim$25\% of the full 0.3--30.0\,\kev\ flux
observed during either epoch. The observed 0.3--30.0\,\kev\ X-ray
luminosities from the two epochs (assuming isotropy and before any absorption
corrections) are $L_{\rm X,1} = (1.90 \pm 0.03) \times 10^{40}$\,\ergps\ and
$L_{\rm X,2} = (3.35 \pm 0.03) \times 10^{40}$\,\ergps, for a distance to
Holmberg\,IX of 3.55\,Mpc (\citealt{Paturel02}). Correcting for the
absorption column inferred from our spectral analysis\footnote{We use
the DISKBB+SIMPL$\otimes$COMPTT model to calculate fluxes, but note that
the models using DISKPBB instead of COMPTT give equivalent fits, and similar
absorption columns are obtained.} (Table \ref{tab_param}), these correspond to
intrinsic 0.3--30.0\,\kev\ luminosities of $L_{\rm int,1} = (2.21 \pm 0.05)
\times 10^{40}$\,\ergps\ and $L_{\rm int,2} = (3.91 \pm 0.08) \times
10^{40}$\,\ergps.

\begin{table}
  \caption{Observed Fluxes for Holmberg\,IX X-1$^{4}$}
\begin{center}
\begin{tabular}{c c c c}
\hline
\hline
\\[-0.25cm]
Epoch & \multicolumn{3}{c}{Fluxes ($10^{-12}$\,\ergpcmsqps)} \\
\\[-0.25cm]
& 0.3--10.0\,\kev\ & 10.0--30.0\,\kev\ & 0.3--30.0\,\kev\ \\
\\[-0.3cm]
\hline
\hline
\\[-0.15cm]
1 & $9.8 \pm 0.1$ & $2.8 \pm 0.1$ & $12.6 \pm 0.2$ \\
\\[-0.2cm]
2 & $19.2 \pm 0.2$ & $3.0 \pm 0.1$ & $22.2 \pm 0.2$ \\
\\[-0.2cm]
\hline
\hline
\\[-0.15cm]
\end{tabular}
\label{tab_flux}
\end{center}
\end{table}

The \nustar\ data confirm that the curvature observed previously in the
3--10\,\kev\ bandpass (\eg \citealt{Stobbart06, Gladstone09, Walton13hoIXfeK})
is a genuine spectral cutoff, as is suggested by the weak \integral\ detection
presented by \cite{Sazonov13}. This is also similar to the results observed from
other extreme ($L_{\rm X} \gtrsim 10^{40}$ \ergps) ULXs observed by \nustar\
to date, \eg Circinus ULX5 (\citealt{Walton13culx}), NGC\,1313 X-1 
(\citealt{Bachetti13}) and IC\,342 X-1 (\citealt{Rana14}, \textit{submitted}).
Indeed, the broadband spectrum of \hoix\ from the first epoch is remarkably
similar to the latter two sources. Neither epoch is consistent with emission from
the standard low/hard state commonly exhibited by Galactic BHBs (see
\citealt{Remillard06rev}), as would broadly have been expected if \hoix\ were an
IMBH accreting at substantially sub-Eddington accretion rates, under the
assumption that the accretion geometry is independent of black hole mass. It
therefore seems likely that we are viewing an unusual high-Eddington phase of
accretion, the physics of which we are only just beginning to probe.

During each epoch, the spectra appear to be best described with a combination
of two thermal components, potentially with a powerlaw-like tail present at the 
highest energies observed, although the contribution of this latter component is
somewhat model dependent. In both epochs, one of these components is
consistent with being a standard thin accretion disk (\citealt{Shakura73}),
although the second (hotter) thermal component deviates from this independent
of the inclusion of a high energy tail. The hotter component can be modeled with
either an optically-thick thermal Comptonization model (which strongly requires
the additional high energy powerlaw-like tail), or a multi-color blackbody disk
model in which the radial temperature profile deviates from that expected
from a simple thin disk. If an additional powerlaw-like emission component truly
is present at the highest energies probed, it generally appears to be very steep
(see Figure \ref{fig_evol}), potentially similar to the very high/steep powerlaw
state exhibited by Galactic BHBs at relatively high accretion rates
(\citealt{Remillard06rev}).

Two component thermal models could plausibly represent a variety of scenarios.
The cooler \diskbb\ could be standard disk emission (see also
\citealt{Miller13ulx}) from the outer accretion flow, while the hotter,
non-standard thermal component could arise from the inner disk where the
effects of radiation pressure are the most prominent, and drive the disk structure
away from the standard thin disk scenario (\eg \citealt{Abram88}). Alternatively,
the soft thermal component could be emission from the photosphere of a
massive optically thick outflow (\eg \citealt{Middleton11a, Sutton13uls}), which
may display some similarity to standard disk emission, with the hotter emission
again arising in the distorted inner disk, broadly similar to the picture for
high/super-Eddington accretion proposed by \cite{Poutanen07}; see also
\cite{Dotan11} and \cite{Ohsuga11}. 

Alternatively, \cite{Dexter12} proposed a framework in which the surface of the
disk is inhomogeneous, resulting in surface temperature fluctuations and thus
deviations in the relative contribution of the emission at different temperatures
from the simple thin disk approximation. Such inhomogeneities could potentially
result in a highly distorted thermal spectrum, and might possibly be able to
simultaneously explain both of the thermal components required to model the
observed spectra, as suggested in the context of ULXs by \cite{Miller13ulx,
Miller14}. As discussed in \cite{Miller14}, such inhomogeneities could arise
naturally through the `photon-bubble' instability (\eg \citealt{Gammie98,
Begelman02}), which may play an important role in high-Eddington accretion
flows. 

Finally, we stress again that while the hotter component has also previously been
interpreted as an optically-thick corona (\eg \citealt{Gladstone09, Pintore14}),
the necessity for an additional Comptonising continuum at higher energies with
such a model renders such a physical interpretation highly unlikely, and favors
the hot, inner accretion disk scenarios outlined above.

\subsection{Spectral Variability}

\begin{figure*}
\epsscale{1.13}
\hspace*{-0.5cm}
\vspace*{0.25cm}
\plotone{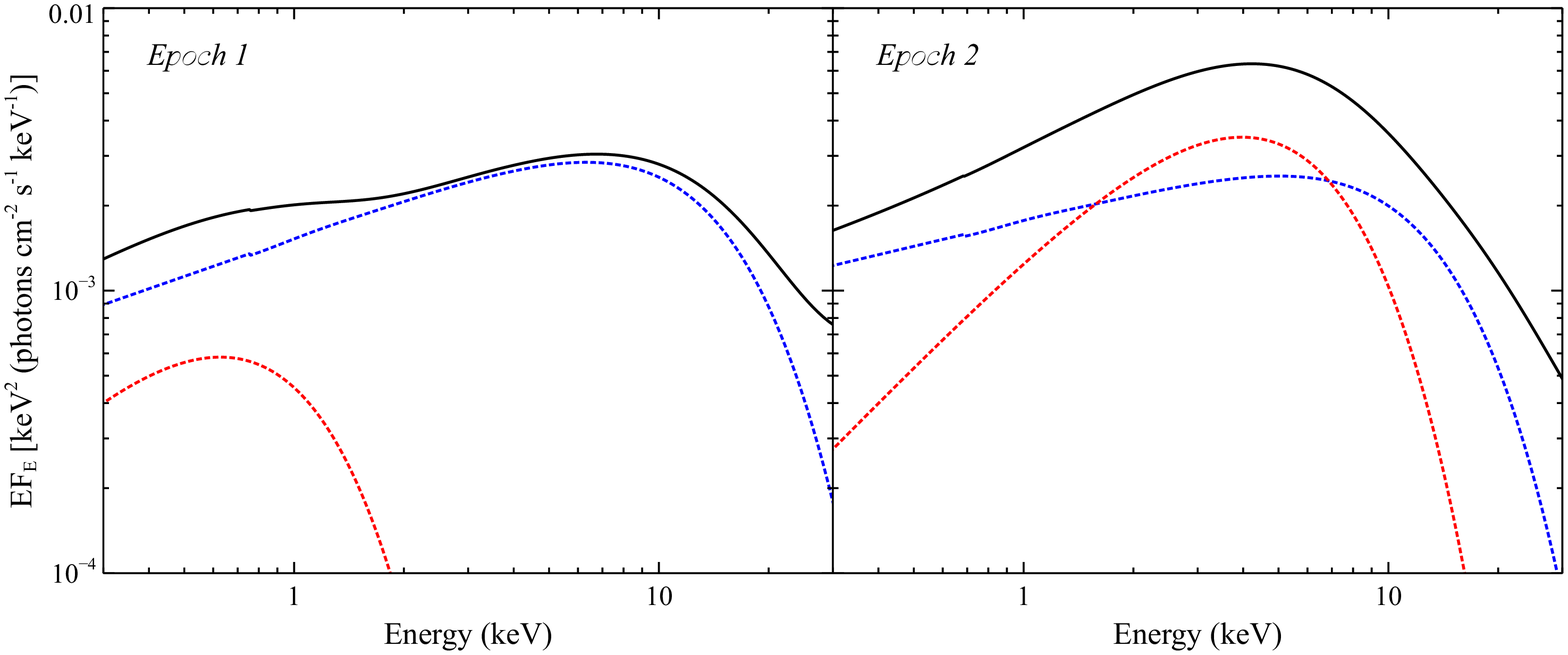}
\hspace*{-0.5cm}
\plotone{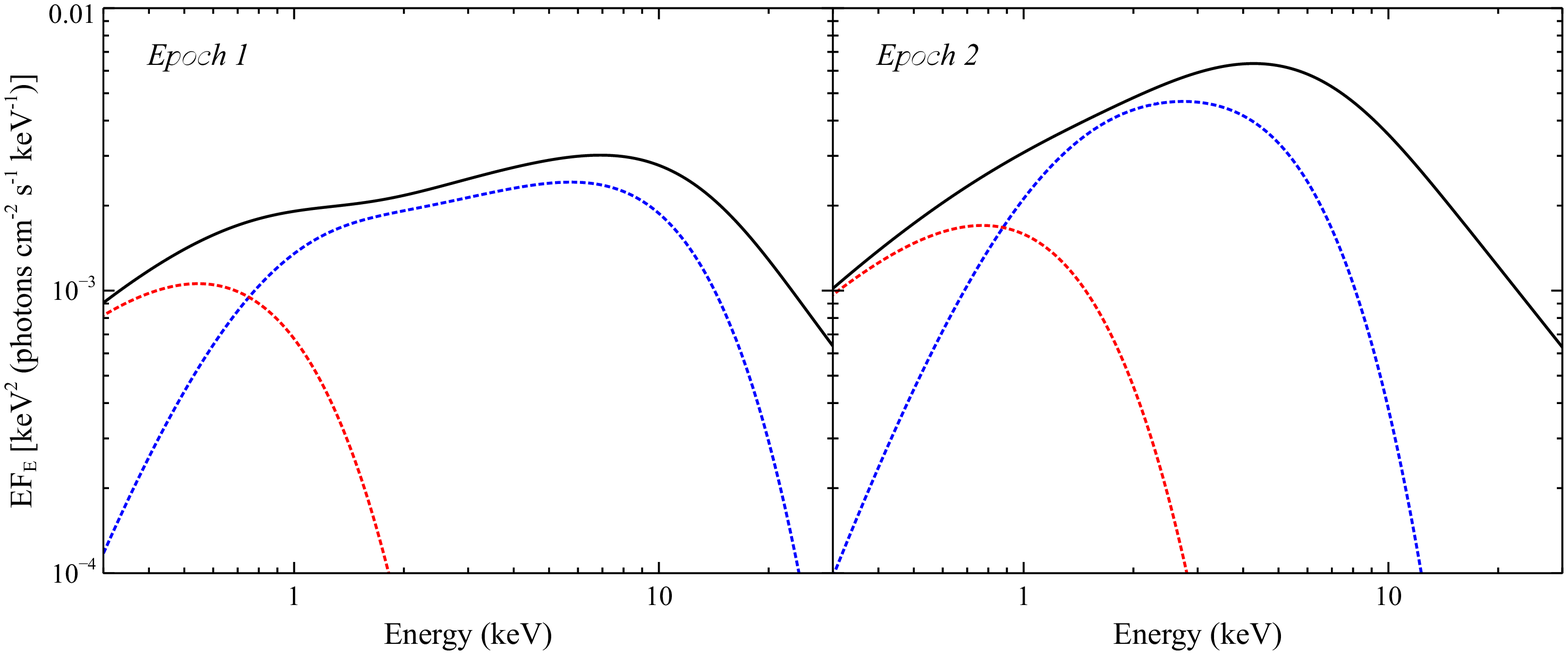}
\caption{
\textit{Top panels:} the relative contribution of the various spectral components
inferred with the DISKBB+SIMPL$\otimes$DISKPBB model (corrected for
absorption) for each epoch. The full model is shown in black, while the DISKBB
contribution is shown in red, and the DISKPBB contribution (before modification
by SIMPL) is shown in blue. In this case, the spectral evolution is dominated by
the cooler of the two thermal components (represented by DISKBB). \textit{Bottom
panels:} The same as the top panels, but for the DISKBB+SIMPL$\otimes$COMPTT
model. Here, the DISKBB contribution is shown in red, and the COMPTT
contribution (before modification by SIMPL) is shown in blue. In this case, the
spectral evolution is dominated by the hotter of the two thermal components
(represented by COMPTT).}
\label{fig_evol}
\end{figure*}

The most striking aspect of these broadband observations is the spectral
variability observed between the two epochs (see Figure \ref{fig_eeuf}). In the
second epoch \hoix\ was observed in a much brigher state, in which both the
temperature and the flux of one of the two thermal components appears to
have changed significantly in comparison to the first. However, as the emission
at the lowest ($E \lesssim 1$\,\kev) and the highest ($E \gtrsim 10$\,\kev)
energies observed has remained remarkably constant in comparison to the
intermediate energies, we find that this evolution can be equally well described
as the low temperature component increasing both its temperature (from
$\sim$0.3 to $\sim$1.5\,\kev) and its flux while the high temperature
component remains broadly stable (case 1), and as the high temperature
component decreasing its temperature (from $\sim$4 to $\sim$1.5\,\kev)
and increasing its flux while the low temperature component remains broadly
stable (case 2). The latter case strongly requires the additional presence
of a high energy powerlaw tail, while the picture is less clear cut for the former,
with the addition of such a component providing a reasonable improvement to
the fit in the first epoch, but only a marginal improvement in the second (see
sections \ref{sec_epoch1} and \ref{sec_epoch2}). These two evolutionary
scenarios are shown in Figure \ref{fig_evol}, represented by the
\diskbb+\simpl$\otimes$\diskpbb\ and \diskbb+\simpl$\otimes$\comptt\
models from each epoch respectively\footnote{Although for the latter model
the statistical requirement for the SIMPL component is not strictly significant
for the second epoch (see section \ref{sec_epoch2}), we show this model for
direct comparison with the first epoch.}.

Case 1 requires the flux and the temperature of the evolving emission
component to be positively correlated, as broadly expected for thermal
emission. However, the observed evolution would have to deviate strongly from
the $L \propto T^{4}$ relation expected for simple blackbody emission with a
constant emitting area, as would be expected for a stable, geometrically thin
disk. Instead, the temperatures and fluxes obtained in case 1 imply the
evolution would have to follow a much flatter, almost linear relation, \ie $L
\propto T$. In contrast, the results for case 2 imply an anti-correlation between
the temperature and the flux of the evolving component, opposite to the basic
expectation for thermal emission, and is therefore challenging to explain.

Considering first case 1 (\ie the lower temperature component dominating the
evolution), our recent analysis of the extreme ULX Circinus ULX5 also found a
luminosity--temperature relation significantly flatter than the basic expectation
for blackbody emission (in that case, $L \propto T^{1.7}$;
\citealt{Walton13culx}), although the deviation from $L \propto T^{4}$ in
Circinus ULX5 was not as extreme as inferred here. As discussed in that work,
a shallow $L-T$ relation would imply either that the inner radius of the disk
decreases with increasing luminosity (geometric changes), or that the color
correction factor (\fcol, which relates the observed surface temperature of the
disk to the effective blackbody temperature via $T_{\rm in} = f_{\rm col}T_{\rm
eff}$) increases with increasing luminosity (atmospheric changes). For Circinus
ULX5, the latter scenario is preferred, due to its high luminosity.

However, in the case of Circinus ULX5, the identification of the thermal
component as emission from the inner disk is likely robust, as the high energy
emission was well explained as Comptonization in an optically thin corona. For
\hoix, this is certainly not the case, and the higher energy emission instead
appears to be dominated by a second thermal emission component in this
scenario (see Figure \ref{fig_evol}), seemingly associated with optically thick
material. Given its higher temperature, it is natural to assume this material
represents the inner disk, and thus that the cooler emission arises from further
out. This association is supported by the failure of models that associate
this emission with the Comptonizing corona; optically-thin coronal emission is
strongly ruled out, and optically-thick coronal emission in turn  ends up
strongly requiring a second, steep Comptonized continuum in the \nustar\
bandpass. Furthermore, recent work by \citealt{Tao13} suggest hot, luminous
accretion disks may naturally produce such tails. 

Given the stability inferred for the high temperature emission in this scenario, a
strong wind may be required to remove the fluctuations observed in the cooler,
more distant regions before they propagate through to the inner flow. In order
to efficiently dampen out these fluctuations, the wind would most likely have to
remove a very large fraction of the accreted mass at the point it is launched,
particularly at higher luminosities. This evolutionary scenario could be
considered broadly similar to certain aspects of those proposed in
\cite{Sutton13uls} and \cite{Pintore14}, who associate the soft component with
emission from a large outflow, and suggest that as the accretion rate increases
the emission from this component should become more prominent, although
as discussed previously this interpretation of the soft emission is not unique
(\citealt{Miller14}). We note that unambiguous signatures of such an outflow in
the form of the ionised iron \ka\ absorption features associated with disk winds
in Galactic binaries are not observed in \hoix\ (\citealt{Walton12ulxFeK,
Walton13hoIXfeK}), suggesting either such outflows are absent, or directed away
from us such that they do not obscure our \los\ to the inner accretion flow.

Recently \cite{Middleton13} investigated atomic features in the iron-L bandpass
($\sim$1\,\kev) for two luminous ULXs with soft X-ray spectra, NGC\,5408 X-1
and NGC\,6946 X-1, suggesting that these features could be explained, at least
in part, by broad absorption features associated with an outflow. On inspection,
there is a tentative suggestion of the presence of residual features at similar
energies here, particularly during the second epoch. However, if real, it is unlikely
they are absorption features in this case, given the lack of accompanying iron
\ka\ absorption features (see above) which are predicted by the models
generated in \cite{Middleton13}, implying an emission origin instead. It is
interesting to note that if the residuals from the second, brighter epoch (see
$\sim$1\,\kev\ in Figure \ref{fig_obs2_ratio}) are interpreted as atomic emission
(\eg emission from a thermal plasma), the strength of the emission must vary
from epoch to epoch, apparently in response to the ULX continuum emission.
Thus, they would have to be associated with the ULX rather than with steady
diffuse emission. However, the exact profile of the residuals from the two epochs
is not always consistent for each of the operational detectors, so any physical
interpretation must be treated with caution, as the residuals may simply relate to
calibration uncertainties. Observations with the high-resolution \textit{Astro-H}
micro-calorimeter (\citealt{ASTROH_tmp}) will be required to robustly address
this issue as there is insufficient signal in the RGS spectra; currently there is no
clear X-ray atomic evidence for the presence of an outflow in \hoix.

For case 2 (the hotter component dominating the evolution), the anti-correlation
between luminosity and temperature is even more counter-intuitive for thermal
blackbody emission than the shallow correlation required for the first scenario.
However, the short term variability behaviour (see section \ref{sec_var}) may offer
some independent support for this evolutionary scenario. In this case, the need
for an additional high energy powerlaw-like tail to the hotter component is very
strong. As the hotter component decreases in temperature, this high energy tail
makes a more substantial contribution to the spectrum in the 0.3--30.0\,\kev\
bandpass, contributing initially only at the very highest energies in the first
epoch, and then dominating the emission above 10\,\kev\ in the second (see
Figure \ref{fig_evol}). Simultaneously, we see evidence for an increase in the
short-term variability above 10\,\kev\ during the second epoch (Figure
\ref{fig_fvar}). 

In Galactic BHBs, short-term variability is generally associated with strong
coronal emission (\eg \citealt{Churazov01, Homan01}); harder states dominated
by the coronal emission generally display stronger variability than
disk--dominated soft states, and in intermediate states the variability is generally
strongest at the energies at which the corona dominates. In contrast,
since the spectral evolution would be dominated by the low-energy component
in case 1, with the high energy emission remaining stable, one might not expect
to see any evolution in the short-term variability properties at high energies in
this scenario. Furthermore, it is interesting to note that the latest
high-Eddington accretion disk models being generated may actually predict a
regime at the highest accretion rates in which the luminosity and temperature
inferred for the disk display an anti-correlation (Bursa et al. 2014, \textit{in
preparation}). If these models are correct, and the spectral evolution displayed by
\hoix\ is truly described by this second scenario, the implication is that \hoix\ is
accreting at a very high accretion rate on the Eddington scale.

Determining the origin and the nature of this spectral evolution is important for
furthering our understanding of the nature of the accretion onto \hoix, and how
the physical structure of the accretion flow might evolve. However, the two broad
evolutionary scenarios allowed by the current broadband observations do not
appear to be obviously distinguished on physical grounds; both require highly
non-standard behaviour, which could nevertheless be considered plausible
under certain circumstances. Further broadband observations probing a more
diverse range of flux states will be required to determine the true origin of the
remarkable evolution displayed by \hoix.

\section{Conclusions}
\label{sec_conc}

We have presented results from the coordinated broadband X-ray observations
of the extreme ULX \hoix\ performed by \nustar, \xmm\ and \suzaku\ in late
2012. The \nustar\ detections provide the first high-quality spectra of \hoix\ at
hard X-rays to date, extending our X-ray coverage up to $\sim$30--35\,\kev.
Observations were undertaken during two epochs, between which \hoix\
exhibited strong spectral variability. Neither epoch is consistent with emission
from the standard low/hard accretion state, as would have been expected if
\hoix\ harbors an IMBH accreting at substantially sub-Eddington rates; the
\nustar\ data confirm in each case that the curvature observed previously in the
3--10\,\kev\ bandpass is a true spectral cutoff. Instead, the spectrum appears
to be dominated by two optically thick thermal components, likely associated
with a distorted accretion disk, with a non-thermal tail also potentially detected
at the highest energies probed. The data allow for either of the two thermal
components to dominate the spectral evolution, although we find that both
scenarios require highly non-standard behavior for behavior for thermal
accretion disk emission. Further broadband observations covering a broader
range of fluxes will be required to determine which component is truly
dominating the observed evolution.

\section*{ACKNOWLEDGEMENTS}

The authors would like to thank the referee for providing useful feedback, which
helped to improve the manuscript. This research has made use of data obtained
with the \mbox{\nustar} mission, a project led by the California Institute of
Technology (Caltech), managed by the Jet Propulsion Laboratory (JPL) and funded
by NASA, \xmm, an ESA science mission with instruments and contributions
directly funded by ESA Member States and NASA, and \suzaku, a collaborative
mission between the space agencies of Japan (JAXA) and the USA (NASA). We
thank the \nustar\ Operations, Software and Calibration teams for support with
the execution and analysis of these observations. This research was supported
under NASA grant No. NNG08FD60C, and has made use of the \nustar\ Data
Analysis Software (\nustardas), jointly developed by the ASI Science Data Center
(ASDC, Italy) and Caltech (USA). We also made use of the NASA/IPAC Extragalactic
Database (NED), which is operated by JPL, Caltech, under contract with NASA.
Many of the figures included in this work have been produced with the Veusz
plotting package: http://home.gna.org/veusz, written and maintained by Jeremy
Sanders. DB and MB are grateful to the Centre National d'Etudes Spatiales (CNES)
for funding their activities.

{\it Facilites:} \facility{NuSTAR}, \facility{XMM}, \facility{Suzaku}

\bibliographystyle{/Users/dwalton/papers/mnras}

\bibliography{/Users/dwalton/papers/references}

\label{lastpage}

\end{document}